# Quantification of Entanglement Entropies for Doubly Excited States in Helium

Chien-Hao Lin and Yew Kam Ho

Institute of Atomic and Molecular Sciences, Academia Sinica, Taipei, Taiwan

March 6th, 2015

## Abstract

In this work, we study the quantum entanglement for doubly excited resonance states in helium by using highly correlated Hylleraas type functions to represent such states of the two-electron system. The doubly-excited resonance states are determined by calculation of density of resonance states under the framework of the stabilization method. The spatial (electron-electron orbital) entanglement measures for the low-lying doubly excited $2s^2$, $2s3s$, and $2p^2$ $^1S^e$ states are carried out. Once a resonance state wave function is obtained, the linear entropy and von Neumann entropy for such a state are quantified using the Schmidt-Slater decomposition method. To check the consistence, linear entropy is also determined by solving analytically the needed four-electron (12-dimensional) integrals.

1. Introduction

   Since the entanglement property plays a crucial role in areas such as quantum teleportation, quantum computation, and quantum cryptography [1], the quantum entanglement in two interacting particles systems have attracted much attention [2]. In particular, in recent years considerable effort has been made on studies of entanglement for two-electron systems including model atoms, quantum dots systems, and natural two-electron atoms. Works on model atoms such as the Moshinsky atom [3, 4, 5, 6], the Crandall atom [7, 8, 9] and the Hooke atom [7, 10, 11] have also been reported in the literatures, as well as works on quantum dot systems [12, 13, 14, 15, 16]. Recently, interest has moved toward the investigation of entanglement in natural two-electron systems, such as the helium atom [7, 8, 17, 18 19, 20, 21]. Dehesa *et al*. [17, 18] explored the helium ground and excited states with Kinoshita-type wave functions and made use of the Monte Carlo multidimensional integration scheme to solve the 12-dimensional integrals needed in calculations of linear entropy. Lin *et al*. have calculated the linear entropy [20] and von Neumann entropy [21] of the helium ground and excited states represented by configuration interaction with *B*-spline basis functions. Benenti *et al*. [22] obtained the linear entropy and von Neumann entropy by employing configuration interaction basis wave functions constructed with Slater type orbital. In our recent works, we studied the linear entropy of the ground state in helium and helium-like atoms, including hydrogen negative ion and positronium ion [23, 24, 25]. Koscik and Okopinska [26, 27] have also reported calculations of entropies for two-electron atomic systems using the Schmidt decomposition method with the original form of Hylleraas wave functions. Quantification of entanglement entropies were also carried out by using Gaussian type basis functions [28, 29]. In the present work, our research is expanded toward doubly excited states. We employ the Hylleraas-type basis to represent the wave functions. As such resonance states are located in the scattering continuum; the usual Rayleigh-Ritz variational bound principal for bound states is no longer valid, and we hence adopt the stabilization method to calculate the density of resonance states [30, 31, 32, 33]. Once the wave function of a resonance state is obtained, it can then be used to calculate linear and von Neumann entropies by using the Schmidt-Slater decomposing method [24, 25]. Furthermore, to check the consistence of our results, we also carry out calculations of linear entropy using the direct integration method that involved four-electron integrals to treat the 12-dimensional integral [23]. By systematically changing the size of our expansion set, we have obtained reasonably accurate results for doubly excited $2s^2$, $2s3s$, and $2p^2$ $^1S^e$ states in the helium atom. Atomic units are used throughout the present work.

2. Theoretical Method

   The non-relativistic Hamiltonian (in atomic units) describing the three-body atomic system, with the nucleus being infinitely heavy, is given by

$$H = -\frac{1}{2}\nabla_1^2 - \frac{1}{2}\nabla_2^2 - \frac{2}{r_1} - \frac{2}{r_2} + \frac{1}{r_{12}} \quad , \tag{1}$$

where 1 and 2 denote the electron 1 and electron 2, respectively, and $r_{12}$ is the relative distance between the electron 1 and electron 2. For *S*-states we use Hylleraas-type wave functions to describe the system, with

$$\Psi_{kmn} = \sum_{kmn} C_{kmn} \{\exp(-\alpha r_1 - \beta r_2) r_1^k r_2^m r_{12}^n + (1 \leftrightarrow 2)\} \quad , \tag{2}$$

where $k + m + n \leq \omega$, and $\omega$, $k$, $m$ and $n$ are positive integers or zero. In the present work we use wave functions up to $N=203$ terms, with $\omega = 11$. The ground state of the atom is of singlet-spin state, denoted as $1s^2\ ^1S^e$. As the spin parts of the atom are antisymmetric, the spatial parts are hence symmetric, as shown in Eq. (2). In the wave functions (Eq. (2)), we take $\alpha = \beta$ to be the nonlinear parameter.

By choosing a set of non-linear parameter $\alpha$, we can plot the stabilization graph. The resonance wave function shows a stable behavior with respect to the change on the non-linear parameter $\alpha$, and it forms a plateau near the resonance energy. From the stabilization plot, we can obtain the density of states by calculating the inverse of the slope,

$$\rho_n = \frac{\alpha_{n+1} - \alpha_{n-1}}{E_{n+1} - E_{n-1}} \quad , \tag{3}$$

where $E_{n+1}$ and $E_{n-1}$ are the energies calculated using the wave functions with the $(n+1)^{th}$ and $(n-1)^{th}$ values, respectively, of the non-linear parameter $\alpha$, as shown in Eq. (2). The resonance energy and the width can then be determined by fitting $\rho_n$ to the Lorentzian profile,

$$\rho_n(E) = \frac{a(\Gamma/2)}{(E - E_r)^2 + \Gamma^2/4} + b \quad , \tag{4}$$

where $E_r$ and $\Gamma$ are the resonance energy and the resonance width respectively. We choose the $\alpha$ with the highest density of states, *ie*, the center of such resonance, to construct the wave function of doubly excited states, and then use the wave function to calculate the linear entropy and von Neumann entropy.

The quantum entanglement of an atomic system can be quantified with entropies, such as von Neumann entropy and linear entropy. The von Neumann entropy of the spatial entanglement for a two-electron system has the form (see [7] for example).

$$S_{vN} = -\text{Tr}(\rho_{red} \log_2 \rho_{red}) \quad , \tag{5}$$

and the linear entropy is defined as

$$S_L = 1 - \mathrm{Tr}\rho_{red}^2 \quad . \tag{6}$$

where $\rho_{red}$ is the one-particle reduced density matrix, and Tr stands for the trace of a matrix. The reduced density matrix can be expressed as

$$\rho_{red}(\mathbf{r}_1, \mathbf{r}_2) = \int \left[\Psi(\mathbf{r}_1, \mathbf{r}')\right]^* \Psi(\mathbf{r}', \mathbf{r}_2) d\mathbf{r}' \quad . \tag{7}$$

To calculate eigenvalues of the reduced density matrix, we adopt the Schmidt-Slater decomposition method. The detail of this computational scheme was presented in Refs. [23 - 25], and here we only point out the highlight of computational procedure in a self-contained manner. A two-electron wave function can be decomposed into a sum of products of one-particle functions by partial wave expansion and as a series of Legendre polynomials,

$$\Psi(\mathbf{r}_1, \mathbf{r}_2) = \sum_{l=0}^{\infty} \frac{f_l(r_1, r_2)}{r_1 r_2} P_l(\cos\theta) \quad , \tag{8}$$

and the coefficients $f_l$ will be used for construction of reduced density matrix. The eigenvalues of such a matrix are then used to deduce linear entropy and von Neumann entropy. In Eq. (8) the infinite sum in $l$ is truncated into a finite sum, for practical purposes, with a maximum value $l_{max}=40$. For a given $l$, with the help of Schmidt decomposition, the function $f_l(r_1, r_2)$ can be decomposed as a sum of products of one-particle wave functions. For a real and symmetric wave function, the function $f_l$ can be expended by the Schmidt decomposition:

$$f_l(r_1, r_2) = \sum_{n=0}^{\infty} \lambda_{nl} u_{nl}(r_1) u_{nl}(r_2) \quad , \tag{9}$$

where $u_{nl}$ is a set of one-particle orthonormal basis, and the $\lambda_{nl}$ can be expressed as an eigenvalue problem in a form integral equation

$$\int_0^{\infty} f_l(r_1, r_2) u_{nl}(r_2) dr_2 = \lambda_{nl} u_{nl}(r_1) \quad . \tag{10}$$

Once the elements of the density matrix for a given partial wave are determined (see [24, 25] for details), eigenvalues $\lambda_{nl}$ can be obtained by diagonalization of the partial wave reduced density matrix. In Refs. [24, 25, 26] it was shown that the relationship between $\Lambda_{nl}$ and $\lambda_{nl}$ is given by

$$\Lambda_{nl} = \left(\frac{4\pi\lambda_{nl}}{2l+1}\right)^2 . \tag{11}$$

Finally, von Neumann entropy for spatial entanglement is then expressed as

$$S_{vN} = -\sum_{nl}(2l+1)\Lambda_{nl}\log_2\Lambda_{nl} \quad , \tag{12}$$

and linear entropy for spatial entanglement as

$$S_L = 1 - \sum_{nl}(2l+1)\Lambda_{nl}^2 \quad . \tag{13}$$

Here, we should mention that we emphasis on the spatial entanglement (the electron-electron orbital entanglement) of the two-electron helium atom. For entanglement due to the spin part, readers are referred to some earlier publications [16, 21, 22, 26] for detailed discussions.

To check the consistence, we also calculate the linear entropy (Eq. (6)) for these states using the four-electron integral method, with

$$\begin{aligned}\mathrm{Tr}\rho_{red}^2 &= \int \rho_{red}^2(\mathbf{r}_1,\mathbf{r}_1)d\mathbf{r}_1 \\ &= \iint \rho_{red}(\mathbf{r}_1,\mathbf{r}_2)\rho_{red}(\mathbf{r}_2,\mathbf{r}_1)d\mathbf{r}_2 d\mathbf{r}_1 \\ &= \iint\iint \Psi^*(\mathbf{r}_1,\mathbf{r}_3)\Psi(\mathbf{r}_2,\mathbf{r}_3)\Psi^*(\mathbf{r}_2,\mathbf{r}_4)\Psi(\mathbf{r}_1,\mathbf{r}_4)d\mathbf{r}_1 d\mathbf{r}_2 d\mathbf{r}_3 d\mathbf{r}_4 .\end{aligned} \tag{14}$$

For the treatment of the needed four-electron integrals when correlated Hylleraas-type wave functions are used, readers are referred to our earlier work [23].

## 3. Calculations and Results

In determining the resonance states with the stabilization method, we iterate the non-linear parameter α in the Hylleraas-type wave functions. As shown in Figure 1, there are several stabilization plateaus for each of doubly-excited resonances states below the $N=2$ thresholds, the $2s$ and $2p$ states, of the He$^+$ ion. For each plateau, we convert it to the density of states $\rho_n$ using Eq. (3) and fit the density curve to a Lorentzian profile, Eq. (4), to determine the resonance energy and width. Three figures below (Figure 2 - 4) show the fitting of density of states to Lorentzian function for the $2s^2$, $2s3s$, and $2p^2$ $^1S^e$ states respectively. Among the multiple stabilization plateaus for a state, the one with the best $r^2$ value (closer to 1.0), implying it is the best fit, is chosen to construct the wave function for the state, and from which we calculate the linear entropy and von Neumann entropy with the Schmidt decomposition method [24, 25], and the four-electron integral method [23]. The results of the three resonance states are listed in Table 1-3, respectively. In these tables, we also compare our present stabilization results with those of earlier results [34, 35, 36] for the resonance energies and widths that were obtained using complex-scaling method [37]. It shows agreements on the resonance parameters are quite good. As for the entanglement entropy for such doubly excited resonance states in the helium atom, we are not aware of any published numerical results in the literature for comparison;

notwithstanding that an investigation on some doubly excited states in helium was reported at a meeting [38].

In Tables 1 to 2, in addition to the results obtained by using the Schmidt-Slate decomposition method, we also show the linear entropy results obtained by using the direct integration method as given in Eqs. (6) and (14). It is observed that up to N=125 terms, results from both the Schmidt decomposition method and the four-electron integration method agree with each other very well, in spite of the fact that they are obtained by using two completely different computational schemes. As it would take considerable computer time and may require multiple precision algorithm to achieve accurate results for calculations using four-electron integrals beyond $N$=125 terms, we only employ more extensive wave functions, up to $\omega$ = 11, $N$=203 terms, in calculations using the Schmidt decomposition method.   Next we summarize our results in the form of a ($S_L$, $S_{vN}$) pair. The entropy pair for the $2s^2$ $^1S^e$ state is determined as (0.4617, 1.378), for the $2p^2$ $^1S^e$ state we have (0.7776, 2.451), and for the $2s3s$ $^1S^e$ state, (0.7704, 2.557).

4.  Summary and Conclusion

We have carried out an investigation of quantum spatial (electron-electron orbital) entanglement on doubly excited resonance states in helium using Hylleraas functions to take into account of the correlation effects. Resonance wave functions are obtained by using the stabilization method, and once such wave functions are obtained the Schmidt-Slater decomposition method are subsequently employed to quantify entanglement entropies, i.e. von Neumann entropy and linear entropy, and our numerical results are first reported in the literature. Furthermore, we believe that the present results for the $2s^2$, $2s3s$ and $2p^2$ $^1S^e$ states in the two-electron helium are quite accurate, and that they can be treated as useful references for future investigations on quantification of entanglement entropies in few-body atomic systems.

**Acknowledgement**

This work was supported by the Ministry of Science and Technology of Taiwan.

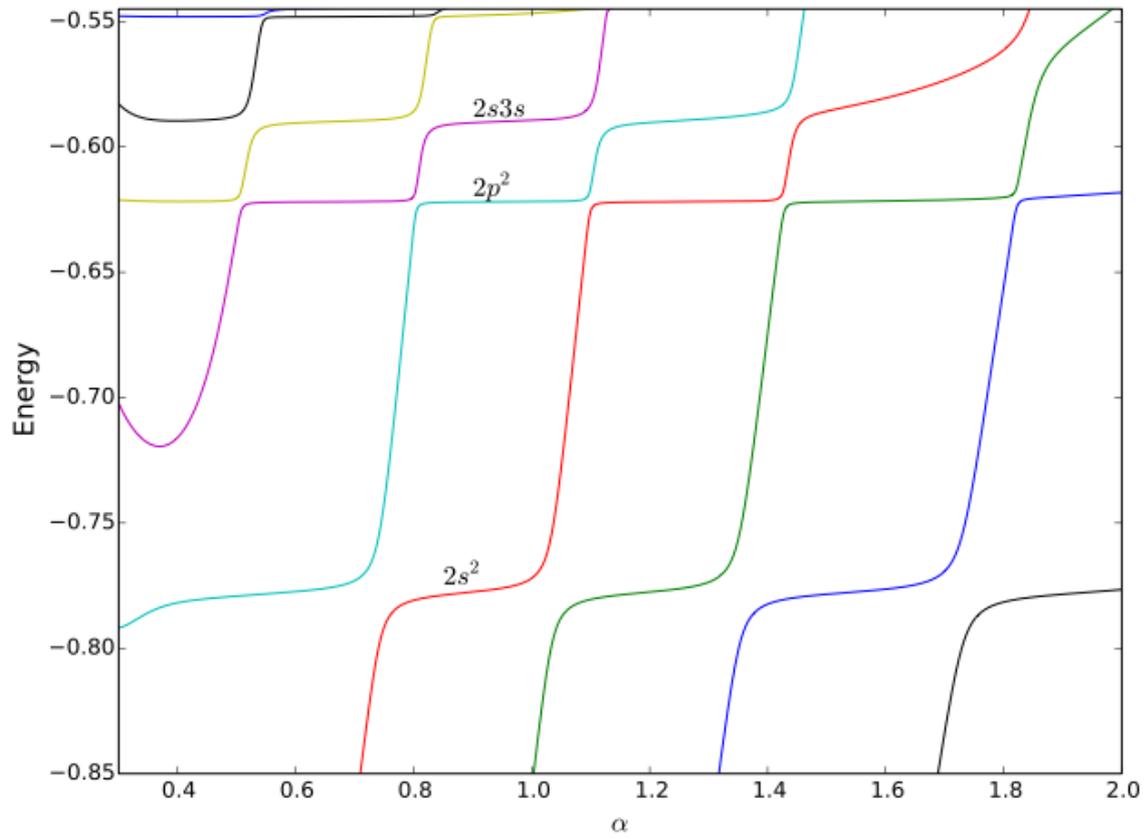

Figure 1. Energy eigenvalues *vs* α parameter for the singlet-spin *S* states of the He atom, with wave functions of *N*=203 terms, $\omega = 11$, showing the $2s^2$, $2p^2$ and $2s3s$ states.

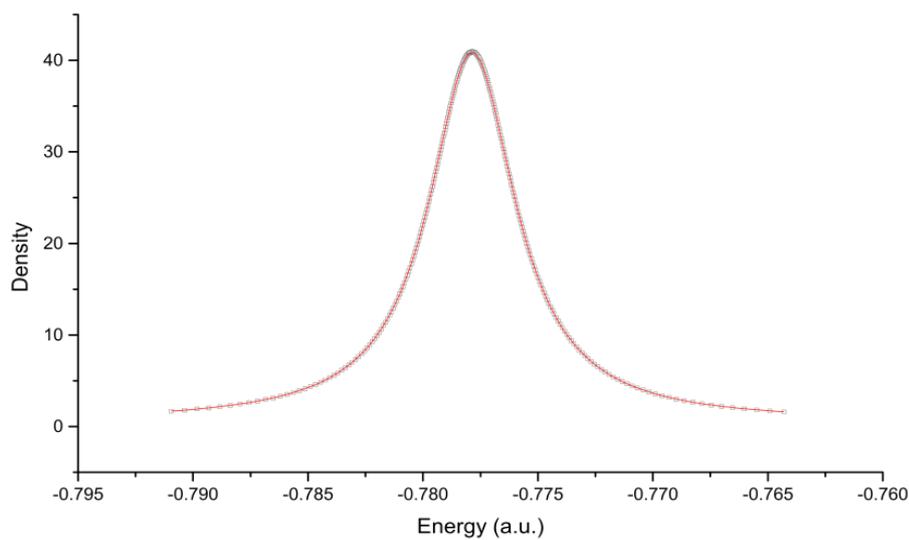

Figure 2. Calculated density (in squares) and the fitted Lorentzian profile (in solid red line) for the $2s^2\ {}^1S^e$ resonance state ($N= 203$ terms)

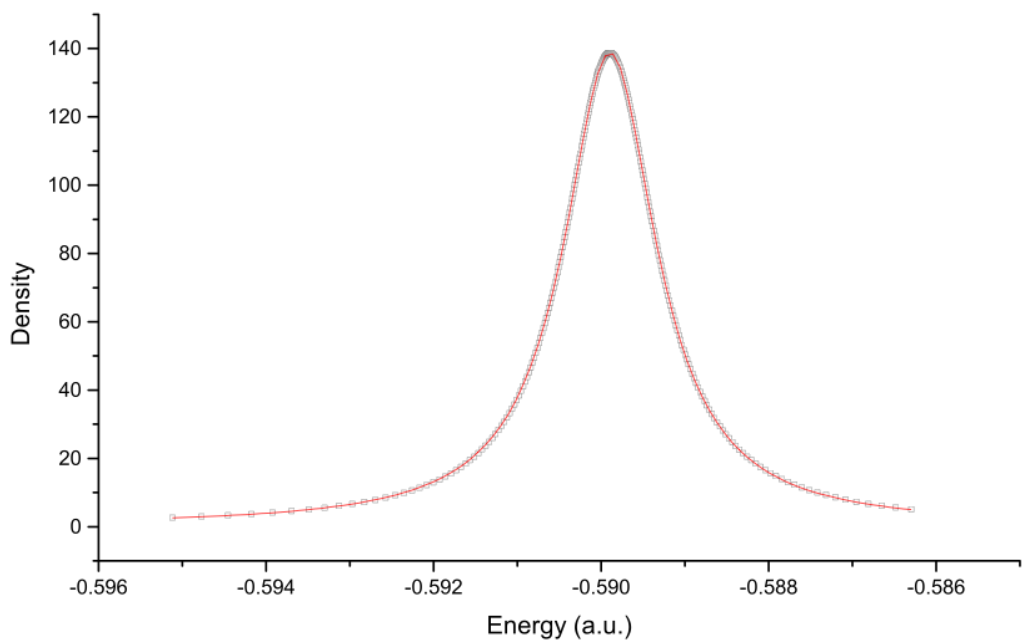

Figure 3. Calculated density (in squares) and the fitted Lorentzian profile (in solid red line) for the $2s3s\ {}^1S^e$ resonance state ($N=203$ terms)

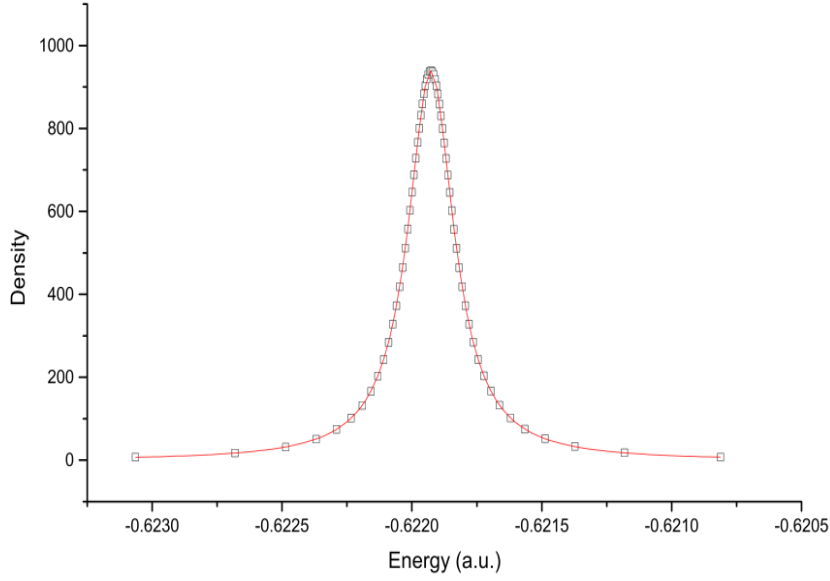

Figure 4. Calculated density (in squares) and the fitted Lorentzian profile (in solid red line) for the $2p^2\ ^1S^e$ resonance state ($N=161$ terms)

Table 1. The calculated energy, width and entropies for the $2s^2\ ^1S^e$ resonance state with different numbers of basis sets.

| N | $E_r$ | $\Gamma$ | $r^2$ | $S_L$ (4-electron) | $S_L$ (Schmidt-Slater) | $S_{vN}$ (Schmidt-Slater) |
|---|---|---|---|---|---|---|
| 70 | -0.7778321 | 0.004523 | 0.99989 | 0.458845 | 0.458846 | 1.359356 |
| 95 | -0.7778694 | 0.004497 | 0.99997 | 0.459458 | 0.459442 | 1.363618 |
| 125 | -0.7778074 | 0.004561 | 0.99993 | 0.459027 | 0.460058 | 1.367720 |
| 161 | -0.7778350 | 0.004564 | 0.99998 | | 0.461261 | 1.375510 |
| 203 | -0.7778583 | 0.004575 | 0.9999996 | | 0.461704 | 1.378501 |
| Other results | -0.777868[a] | 0.00453[a] | | | | |
| | -0.777867[b] | 0.004541[b] | | | | |

(a) Refs. [34, 35];   (b) Ref. [36]

Table 2. The calculated energy, width and entropies for the $2p^2$ $^1S^e$ resonance states with different numbers of basis sets.

| N | $E_r$ | $\Gamma$ | $r^2$ | $S_L$ (4-electron) | $S_L$ (Schmidt-Slater) | $S_{vN}$ (Schmidt-Slater) |
|---|---|---|---|---|---|---|
| 70 | -0.6219657 | 0.0002429 | 0.970444 | 0.777030 | 0.777031 | 2.447453 |
| 95 | -0.6219261 | 0.0002260 | 0.999197 | 0.777616 | 0.777617 | 2.449617 |
| 125 | -0.6219257 | 0.0002149 | 0.999987 | 0.777628 | 0.777630 | 2.449901 |
| 161 | -0.6219270 | 0.0002154 | 0.9999993 | | 0.777614 | 2.449946 |
| 203 | -0.6219259 | 0.0002169 | 0.999976 | | 0.777653 | 2.450665 |
| Other results | -0.6219275[a] -0.6219273[b] | 0.0002156[a] 0.0002156[b] | | | | |

(a) Refs. [34, 35]; (b) Ref. [36]

Table 3. The calculated energy, width and entropies for the $2s3s$ $^1S^e$ resonance states with different numbers of basis sets.

| N | $E_r$ | $\Gamma$ | $r^2$ | $S_L$ (Schmidt-Slater) | $S_{vN}$ (Schmidt-Slater) |
|---|---|---|---|---|---|
| 161 | -0.5898955 | 0.001343 | 0.999932 | 0.770336 | 2.556674 |
| 203 | -0.5898947 | 0.001348 | 0.999990 | 0.770376 | 2.557395 |
| Other results | -0.589895[a] -0.5898946[b] | 0.00135[a] 0.001362[b] | | | |

(a) Refs. [34, 35]; (b) Ref. [36]